\documentclass[12pt]{article}
\pdfoutput=1
\usepackage[colorlinks=true,linkcolor=blue,urlcolor=blue,
  citecolor=blue]{hyperref}
\usepackage{fullpage}
\usepackage{amssymb,graphicx}
\usepackage{amsmath}
\usepackage[margin=0.5cm]{caption}
\usepackage{hyperref}
\usepackage{cite}
\usepackage{upgreek}

\def\d{\partial}
\def\l{\left(}
\def\r{\right)}

\newcommand{\be}{\begin{equation}}
\newcommand{\ee}{\end{equation}}
\newcommand{\bea}{\begin{eqnarray}}
\newcommand{\eea}{\end{eqnarray}}
\newcommand{\bg}{\begin{gather}}
\newcommand{\eg}{\end{gather}}
\newcommand{\bseq}{\begin{subequations}}
\newcommand{\eseq}{\end{subequations}}

\newcommand{\comment}[1]{}

\begin{document}

\baselineskip=15.5pt
\begin{titlepage}
\begin{center}
{\Large\bf Living with the Wrong Sign}
 \\
\vspace{0.5cm}
{ \large
Patrick Cooper, Sergei Dubovsky,  and Ali Mohsen
}\\
\vspace{.45cm}
{\small  \textit{    Center for Cosmology and Particle Physics,\\ Department of Physics,
      New York University\\
      New York, NY, 10003, USA}}\\
\end{center}

\begin{abstract}
    We describe a UV complete asymptotically fragile Lorentz-invariant theory exhibiting
    superluminal signal propagation.  Its low energy effective action contains  ``wrong" sign higher
    dimensional operators. Nevertheless, the theory gives rise to an S-matrix, which is defined at
    all energies. As expected for a non-local theory,
    the corresponding scattering amplitudes  are not exponentially bounded on the
    physical sheet,  but otherwise are healthy.  We study some
    of the physical consequences of this S-matrix.

\end{abstract}
\end{titlepage}


\section{\label{sec:level1}Introduction}
A number of interesting apparently consistent low energy effective theories giving rise to long
distance modifications of gravity has been constructed over the past decade
\cite{Dvali:2000hr,ArkaniHamed:2003uy,Dubovsky:2004sg,Nicolis:2008in,deRham:2010kj}. Explaining the
observed acceleration of the Universe following this path, which was one of the major motivations
for these developments, remains a long shot. Nevertheless, these models are of definite interest
from both theoretical and  phenomenological points of view. In particular, they give rise to a
number of striking observational signatures, including anomalous precession of the Moon perihelion
\cite{Dvali:2002vf,Lue:2002sw}, a strong monochromatic gravitational line from a massive graviton
\cite{Dubovsky:2004ud}, hairy supermassive black holes \cite{Dubovsky:2007zi}, an exotic cosmic
microwave background $B$-mode spectrum \cite{Dubovsky:2009xk}, or a violation of the equivalence
principle for compact astrophysical objects \cite{Hui:2012jb}.

On the theoretical side, all these models are effective field theories with a very low cutoff scale
and it remains to be seen whether any of them can be UV completed into a microscopic theory with
acceptable physical properties.  There is a common underlying physical reason why finding a UV
completion for these models is hard.  One of the prices (or gains, depending on a viewpoint) to pay
for affecting gravity at long distance scales is the possibility of superluminal signal
propagation.  This results in a tension between locality and causality making it impossible to
construct a conventional UV completion \cite{Adams:2006sv}.  The tension is especially severe in
long distance modifications of gravity possessing a Poincar\'e invariant ground state, as required for
applying the arguments of \cite{Adams:2006sv}.  A somewhat related tension in Lorentz-violating
scenarios arises in the presence of black hole horizons \cite{Dubovsky:2006vk}.

The simplest toy model, where the arguments of \cite{Adams:2006sv} apply, describes a single
Goldstone boson $X$ with an effective Lagrangian of the form
\be
\label{simplest}
{\cal L}={1\over 2}(\d X)^2+{c\over \Lambda^4}(\d X)^4+\dots\;.
\ee
It is straightforward to construct a UV completion for this model if the constant $c$ is positive
(we work in the mostly ``$-$" signature). On the other hand, despite its benign appearance, this
simple effective field theory cannot be UV completed into a consistent local microscopic quantum
theory for negative values of $c$. The physical reason for this is the existence of classical
backgrounds (such as $X\propto t$)  exhibiting superluminal propagation of small perturbations.

At this stage  two  viewpoints are possible. The conservative conclusion is that this problem
signals an incurable pathology and effective field theories with wrong sign operators should be
discarded. However, an adventurous person may argue that superluminality simply indicates that one
should look for a UV completion  beyond the realm of conventional quantum field theories. Of course,
to justify the second viewpoint one needs to construct an example of such a UV completion, so that
it becomes possible to judge whether its physical properties are acceptable. The goal of this note
is to provide a first example of this kind and to study its physical properties.

A disclaimer is in order. All of our construction operates in two space-time dimensions.
Two-dimensional theories are special in many respects. In particular, unlike in higher dimensional
theories, in two dimensions superluminality does not allow one to construct smooth backgrounds
exhibiting closed time-like curves. Somewhat related to this, in two dimensions it is possible to
introduce an alternative causal structure, which is compatible with Lorentz symmetry, yet allows for
superluminal (in fact, instantaneous) signal propagation.  Based on this observation examples of UV
complete Lorentz invariant superluminal theories have already been constructed in the past
\cite{Dubovsky:2008bd,Dubovsky:2011ii}. However, the setup considered here relies on peculiarities
of two-dimensional physics to a much lesser extent and the possibility of its extension to higher
dimensional models appears more likely.

\section{Description of the setup}
Our construction is based on recent progress achieved in understanding the dynamics of long
strings \cite{Dubovsky:2012sh,Dubovsky:2012wk}.  One starts with a classical Nambu--Goto action,
describing $(D-2)$ scalar fields $X^i$ (transverse perturbations of a string),
\be
\label{action}
S_{NG} = -\ell_s^{-2}\int d^2\sigma  \sqrt{ - {\mathrm{det}} \l\eta_{a b} - \ell_s^2 \partial_a X^{i}
    \partial_b X^{i}\r}\;,
\ee
where $\ell_s^{-2}$ is the string tension. At the quantum level this may be considered as an
effective theory similar to the Goldstone Lagrangian (\ref{simplest}).  This theory is naively
non-renormalizable. Famously, however, it can be consistently quantized when the number of flavors is
equal to 24, giving rise to a critical bosonic string \cite{Goddard:1973qh}.

The traditional treatment of string dynamics focuses on the properties of short (open or closed)
strings, which is natural if the goal is to provide a target space-time interpretation.  However,
for the purpose of this paper we are mainly interested in a purely two-dimensional interpretation
of (\ref{action}). From this perspective the most natural and simplest set of observables to
consider are the $S$-matrix elements for the scattering of worldsheet perturbations of an infinitely
long string. Geometrically these represent wiggles (or phonons) propagating along the string.  As
explained in  \cite{Dubovsky:2012wk}, in this language the critical string can be {\it defined} by
its exact $S$-matrix. The theory is integrable and reflectionless, so that the two-particle
scattering is purely elastic and is characterized by the phase shift of the form
\be
\label{eps}
 e^{2 i \delta(s)} = e^{i s \ell_s^2/4} ,
\ee
where $s$ is the Mandelstam variable. Phase shifts for multiparticle processes are obtained by
summing pairwise the phase shifts of all the colliding particles.

Despite its simplicity, the phase shift (\ref{eps}) exhibits a number of remarkable properties, and
corresponds to an integrable theory of gravity rather than to a conventional field theory.  In
particular, the UV asymptotics of this theory are not controlled by a conventional fixed point. This
is clear from the form of the phase shift (\ref{eps}) which retains a non-trivial dependence on the
microscopic length scale $\ell_s$ at arbitrarily high energies. Clearly, this is incompatible with
scale invariant behavior in the UV. This gives rise to a number of unconventional features, and this
type of asymptotic behavior was called ``asymptotic fragility".

An asymptotically fragile phase shift (\ref{eps}) satisfies all the properties expected from a
healthy $S$-matrix. In particular, it is analytic and polynomially bounded everywhere {\it on the
physical sheet}, which is usually taken as a definition of locality in the $S$-matrix language. On
the other hand, it exhibits an essential singularity at $s\to \infty$, which prevents one from
defining local off-shell observables, corresponding to (\ref{eps}). This is one of the reasons why
this theory is gravitational, rather than a conventional field theory. Another gravitational
property  of the model is the universal time delay proportional to the center of mass energy of the
collision,
\[
\delta t_{del}=\frac{1}{2}
E \ell_s^2\;,\;
\]
 which may be considered as an integrable precursor of black hole formation and evaporation.

Perhaps the most direct proof that the phase shift (\ref{eps}) indeed defines a critical string
comes from calculating the finite volume spectrum of the theory using the Thermodynamic Bethe Ansatz
technique \cite{Zamolodchikov:1989cf,Dorey:1996re}. This way one exactly reproduces the spectrum of
a critical bosonic string.

As a further check, it is straightforward to see that the phase shift (\ref{eps}) agrees with the
perturbative tree-level and one-loop amplitudes following from the Nambu--Goto action (\ref{action})
at $D=26$. However, the action (\ref{action}) is non-renormalizable and presumably at higher loop
order has to be supplemented with an infinite set of scheme-dependent counterterms to reproduce the
phase shift (\ref{eps}).

Note, that the phase shift (\ref{eps}) defines a consistent relativistic two-dimensional theory for
an arbitrary number of flavors. However, for $D\neq 26$ the one-loop amplitude following from the
Nambu--Goto action differs\footnote{$D=3$ is another interesting exceptional case, c.f.
\cite{Mezincescu:2010yp}.} from (\ref{eps}) by a rational annihilation term (the
Polchinski--Strominger interaction \cite{Polchinski:1991ax}). To reproduce (\ref{eps}) for $D\neq
26$ the action (\ref{action})  needs to be supplemented with a counterterm, which cancels this
effect. This counterterm is perfectly consistent from two-dimensional perspective, but incompatible
with non-linearly realized target space Poincar\'e symmetry.

A further discussion of this family of integrable models can be found in
\cite{Dubovsky:2012sh,Dubovsky:2012wk}. Here, instead, let us move directly to the main point and
describe the superluminal setup.

The basic idea is very simple and motivated by the following observation. Note that, unlike what one
may be used to in more sophisticated quantum theories, the perturbative low energy expansion for the
phase shift (\ref{eps}) is not asymptotic. It has an infinite radius of convergence and the phase
shift is given simply by the sum of all perturbative terms. This suggests that flipping the sign of
the coupling constant, $\ell_s^2$, may also result in a well-defined theory.

So, inspired by the success of this approach for a worldsheet theory of a bosonic string, let us try
to define a new integrable theory by its exact scattering phase shift $\delta_n$ of the form
\be
\label{epsn}
 e^{2 i \delta_n(s)} = e^{-i s\ell_s^2/4} ,
\ee
where, as before, we assume $\ell_s^2>0$.  When expanded at low energies this scattering amplitude
violates the positivity condition of \cite{Adams:2006sv}.  It can be reproduced from the action
(\ref{action}), with $\ell_s^2\to-\ell_s^2$ , and supplemented with the same set of scheme-dependent
counterterms as required to reproduce the conventional phase shift (\ref{eps}). Geometrically this
action describes an infinitely long string in a space-time with $(D-1)$ time coordinates and a
single spatial coordinate. The target space interpretation of such a system is highly obscure, but
for our purposes we do not need it and will consider the system from a purely two-dimensional point
of view.

In agreement with general arguments of \cite{Adams:2006sv} this theory is even more non-local than
the worldsheet theory of a conventional string (\ref{eps}). Namely, not only does the phase shift
(\ref{epsn}) exhibit an essential singularity at the infinity, it is also polynomially
unbounded on the physical sheet.  This represents an interesting step forward compared to the usual
situation with superluminal effective theories. Conventionally, these theories are
non-renormalizable, and one simply concludes that superluminality indicates the absence of the UV
completion. Here, the theory is non-renormalizable, but we managed to construct finite on-shell
scattering amplitudes. We still are not able to determine local off-shell observables, but the same
situation holds for the conventional string worldsheet theory, which is known nevertheless to be an
interesting and healthy physical system. So an interesting question arises to compare the two
models and to characterize in what sense the ``wrong" sign theory is pathological as compared to its
``right" sign cousin, if this is really the case.

In the rest of the paper we will make several steps in this direction, by probing some of the
physics following from the superluminal phase shift (\ref{epsn}). At the technical level all of our
calculations are close counterparts of the corresponding steps in the analysis of the ``right" sign
theory (\ref{eps}). Essentially, they can be performed by flipping the sign of $\ell_s^2$ in the
formulas of \cite{Dubovsky:2012wk}.  Nevertheless, as we will see, the resulting physics turns out
to be quite different.

\section{Classical Time Advance}
Let us start by illustrating superluminal properties of the phase shift (\ref{epsn}) as seen in
the classical limit. For simplicity, let us consider a single flavor case, so that the corresponding
classical action is
\be
\label{action1}
S=\ell_s^{-2}\int d\tau d\sigma\sqrt{1+\ell_s^2(\d X)^2}\;.
\ee
In agreement with the discussion in the {\it Introduction}, superluminality is manifest when
considering a classical background of the form
\[
X_{cl}={v \tau\over \ell_s}\;.
\]
Namely, the quadratic action for  small perturbations $\pi$ around this background takes the form
\[
S_2=\int d\tau d\sigma\l {1\over
2(1+v^2)^{1/2}}(\d\pi)^2-{v^2\over2 (1+v^2)^{3/2}}\l\d_\tau\pi\r^2\r.
\]
This gives rise to a linear dispersion relation
\[
\omega=c_s k
\]
with a superluminal velocity
\[
c_s=\sqrt{1+v^2}\;.
\]
Related to this superluminality the phase shift (\ref{epsn}) gives rise to a time {\it advance}
\be
\label{tadv}
\delta t_{ad}={1\over 2}E\ell_s^2
\ee
for scattering processes around the trivial background $X=0$. It is instructive to see how this time
advance comes out at the classical level. For this purpose let us consider a purely left-moving
field configuration, which is always a classical solution for the action (\ref{action1}),
\[
X_{cl}=X(\tau+\sigma)\;.
\]
Understanding the scattering of a small right-moving perturbation off this background amounts to
studying the null geodesics propagating in the metric induced by the classical solution
\begin{eqnarray}
    ds^2 = \left( \eta_{a b}+\ell_s^2 \partial_a X \partial_b X \right)
    d\sigma^a d\sigma^b \nonumber = (1+\ell_s^2X'^2){d \tau}^2 + 2 \ell_s^2X'^2{d \sigma}
   {d \tau} - (1 -\ell_s^2 X'^2) {d \sigma}^2 \nonumber .
\end{eqnarray}
The null geodesic equation is given by
\begin{eqnarray}
    {d\tau\over d\sigma} = \frac{-\ell_s^2 X'^2 + 1}{\ell_s^2 X'^2 \pm 1} \nonumber ,
\end{eqnarray}
where the upper sign corresponds to a right moving excitation which experiences a non-trivial time
shift. We see this is a time advance since $\tau'< 1$.  Note that $\tau'$ can even become negative
for large enough $X'$.  In this case the right-mover moves ``back in time" at intermediate times,
which simply indicates that $\tau$ is not a good Cauchy time for such a background.  The time
advance is given by
\begin{eqnarray}
    \delta t_{ad} &=& \int_{-\infty}^{\infty} d\sigma (1 - \tau') = \int_{-\infty}^{\infty} d\sigma \frac{2
    \ell_s^2X'^2}{\ell_s^2X'^2 + 1} \nonumber \\
    \label{labadv}
    &=& \int_{-\infty}^{\infty} d\sigma_+ \; \ell_s^2X'^2 (\sigma_+) = \ell_s^2 \Delta E ,
\end{eqnarray}
where $\Delta E$ is the energy of the classical solution relative to the vacuum. This classical
time advance exactly agrees with (\ref{tadv}), following from the exact quantum $S$-matrix, after
one takes into account that the expression (\ref{tadv}) calculates the time shift in the rest frame
of the colliding wave packets, as opposed to the ``lab frame" time advance (\ref{labadv}).
\section{Thermodynamics and Finite Volume Spectrum}
An important piece of physical information, which becomes accessible when the exact $S$-matrix is
known, is the finite temperature equation of state of the system, {\it i.e.} the free energy density
as a function of temperature, $f(T)$. For an integrable theory it can be extracted using the (ground
state) Thermodynamic Bethe Ansatz (TBA) \cite{Zamolodchikov:1989cf}. Moreover, for a Lorentz
invariant theory, free energy determines also the vacuum Casimir energy $E_0(R)$ on a circle of
circumference $R$ through
\begin{eqnarray}
    E_0(R) = R f(R^{-1}) \nonumber\;.
\end{eqnarray}
The derivation of the ground state TBA equations for a superluminal phase shift (\ref{epsn}) is
completely parallel to the one presented in \cite{Dubovsky:2012wk} for a conventional string. As
expected, the result is different only by a flip of a sign in front of $\ell_s^2$.  Namely, the free
energy density is given by
\begin{eqnarray}
    f(R^{-1}) = - {\ell_s^{-2}} + \frac{1}{2 \pi R} \sum_{j=1}^{D-2} \int_0^{\infty} d p'
    \mathrm{ln}\left(1 - e^{-R \epsilon_L^j (p')} \right)
    + \frac{1}{2 \pi R} \sum_{j=1}^{D-2} \int_0^{\infty} d p'
    \mathrm{ln}\left(1 - e^{-R \epsilon_R^j (p')} \right) \nonumber ,
\end{eqnarray}
where the ``pseudoenergies" $\epsilon_{L,R}$ are determined from a system of integral equations of
the form
\begin{eqnarray}
    \epsilon_L^i (p) = p \left( 1 - \frac{\ell_s^2}{2 \pi R} \sum_{j=1}^{D-2} \int_0^{\infty}
    d p' \mathrm{ln} \left(1 - e^{-R \epsilon_R^j (p')}\right) \right) \nonumber \\
    \epsilon_R^i (p) = p \left( 1 - \frac{\ell_s^2}{2 \pi R} \sum_{j=1}^{D-2} \int_0^{\infty}
    d p' \mathrm{ln} \left(1 - e^{-R \epsilon_L^j (p')}\right) \right) \nonumber .
\end{eqnarray}
Just like for a conventional string, these equations are straightforward to solve analytically.
By picking up the solution which approaches the free theory in the limit $\ell_s^2\to 0$, we obtain
\begin{eqnarray}
    E_0(R) = {R}f(R^{-1}) = -\sqrt{\frac{R^2}{l_s^4} + \frac{4 \pi}{l_s^2} \frac{D-2}{12}}
    \nonumber .
\end{eqnarray}
Not surprisingly, this expression closely resembles the ground state energy of a bosonic string.
One major difference however, is that the free energy is real at all temperatures. This means that
the Hagedorn behavior is not present in this theory.  In a sense, the thermodynamics of a
superluminal theory is less pathological than that of a conventional string.

Related to this  the dependence of the energy density on the pressure,
\begin{eqnarray}
    \label{eos}
    \rho = \frac{p}{1+l_s^2 p} \;,
\end{eqnarray}
does not exhibit a singularity, which was present for a conventional string.

The absence of the Hagedorn behavior is straightforward to understand. From a two dimensional
perspective the Hagedorn behavior arises as a consequence of a large binding energy following from the
phase shift (\ref{eps}).  This results in the fast growth of the density of states implying the
divergence of the heat capacity in the thermodynamic limit of the theory. By flipping the sign of
$\ell_s^2$ we replaced attraction with repulsion, which eliminated the Hagedorn behavior.

To confirm this interpretation let us take a look at the spectrum of the excited states of the
superluminal theory in a finite volume.  Following the derivation in \cite{Dubovsky:2012wk} we
obtain
\begin{eqnarray}
    \label{enspec}
    E(N,\tilde{N}) = -\left( \frac{4 \pi^2 (N - \tilde{N})^2}{R^2} + \frac{R^2}{l_s^4}
    -\frac{4 \pi}{l_s^2} \left( N + \tilde{N} - \frac{D-2}{12} \right)
    \right)^{1/2} \;,
\end{eqnarray}
where the positive integers $N,\tilde{N}$ count the total left- and right-moving KK momentum of a
state (in units of $2\pi/R$).  For a conventional bosonic string these would be called the levels of
a state.  It is straightforward to see now that the two-particle binding energy is positive
\[
\Delta E= E(1,1)-E(0,0)-2(E(1,0)-E(0,0))>0
\]
{\it i.e.}, the interaction is repulsive.

\begin{figure}
    \begin{center}
    \includegraphics[width = 12 cm]{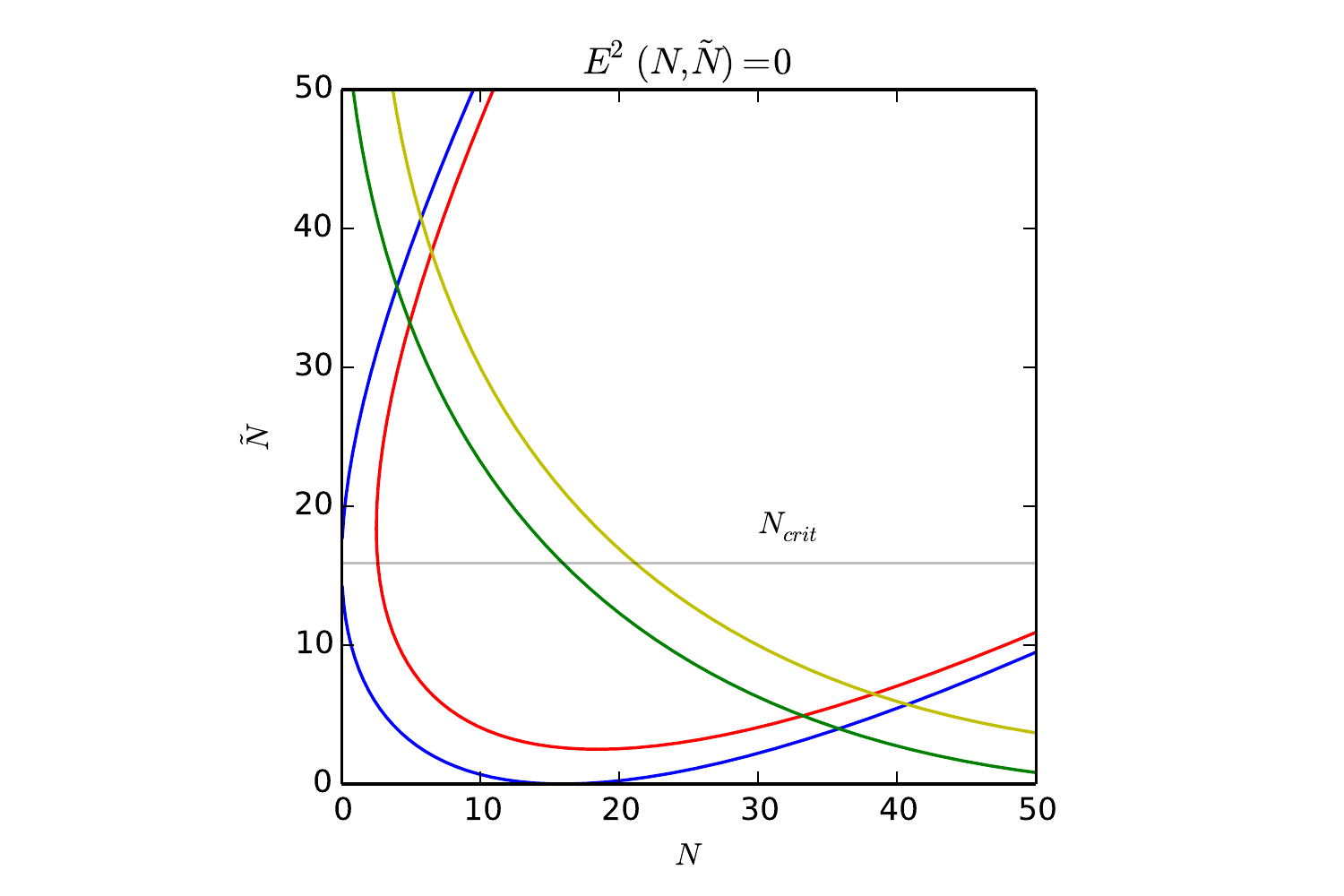}
    \caption{Above is a plot of the region where energy becomes imaginary.  Blue and green are at $D
    = 2$, $R/l_s = 10$ and $20$ respectively. Red and Yellow are at D = 62, $R/l_s = 10$ and
    $20$ respectively. The allowed region is the left of each curve.
\label{Nplane}}
\end{center}
\end{figure}

A further inspection of the spectrum (\ref{enspec}) reveals, however, that in spite of the absence
of the Hagedorn behavior for the ground state, this spectrum exhibits an even more bizarre
property. Namely, for {\it any} value of $R$ there are infinitely many states with imaginary
energies. We illustrated this behavior in Fig.\ref{Nplane}, where we have shown the region of
positive and negative $E^2$ in the $(N,\tilde{N})$ plane for several values of $D$ and $R/\ell_s$.
This property indicate that the superluminal theory cannot be put in a finite volume, at least in a
conventional way.

Pathological at first sight, this behavior is straightforward to understand. Indeed, as a
consequence of the time {\it advance} (\ref{labadv}) two points at the same constant time surface
become causally connected if the ``string" between them is excited to sufficiently high energy.
Identification of these two points is clearly a bad idea then, because it results in a closed
time-like curve. From (\ref{labadv}) we can estimate that this happen when
\[
\ell_s^2\Delta E\simeq {2\pi\ell_s^2 N\over R}
\]
becomes of order $R$. This nicely matches what we observe in Fig.\ref{Nplane}, namely the $E^2=0$
curve with $D=2$ (which corresponds to the classical answer for the spectrum) touches the $N$-axis
at
\[
N_{crit}={R^2\over 2\pi\ell_s^2}\;.
\]
We see that the proliferation of states with imaginary energies in the finite volume cannot be
considered as a pathology of the theory, but rather is a sign that we attempted to perform an
illegal action.  It would be interesting to find whether the model can nevertheless be consistently
put in a finite volume. Classically, this would require identification of points on Cauchy
slices, which are background dependent and in general do not correspond to constant time surfaces.
TBA, at least in its standard form, is not an appropriate tool in such a situation, so currently we
do not know how to proceed in the quantum theory. It is also puzzling that the TBA result exhibits
an island of apparently healthy $E^2>0$ states with arbitrary large $N>N_{crit}$ (see
Fig.\ref{Nplane}). Classically these contain a closed time-like curve, so that their physical
interpretation is obscure.

\section{Cosmology}
Another interesting property of the conventional worldsheet theory is the presence of cosmological
backgrounds \cite{Dubovsky:2012wk}.  Let us see what these look like in the superluminal theory. For
simplicity, we restrict to a single field case.  In a $(1+1)$-dimensional world isotropy is not a
constraint, so one is looking for homogeneous solutions. Imposing an invariance under coordinate
translations, $\sigma\to \sigma +const$ leaves one with only trivial vacuum solutions. Instead,
following \cite{Dubovsky:2012wk}, one can look for homogeneous solutions where the isometry is
generated by boosts.  In fact, there is a larger set of solutions, labeled by a continuous parameter
$\gamma$, such that the isometry is generated by the combination of a boost and a shift of the field
of the form
\be
\label{isometry}
X(\sigma^{+},\sigma^-)\to X((1+\epsilon)\sigma^+,(1+\epsilon)^{-1}\sigma^-)-\epsilon \gamma\;,
\ee
where $\sigma^\pm=\tau\pm\sigma$.
A general field configuration invariant under (\ref{isometry}) can be presented in the form
\be
\label{Ansatz}
\ell_sX(\sigma^+\sigma^-)=f(\sigma^+\sigma^-)+{\gamma\over 2}\log{\sigma^+\over\sigma^-}\;.
\ee
Then the field equation becomes
\be
\d_{\alpha^2}\l{\alpha^2\d_{\alpha^2}f\over \sqrt{1+\alpha^2 (\d_{\alpha^2}f)^2-{\gamma^2\over
4\alpha^2}} } \r=0\;,
\ee
where $\alpha^2=\sigma^+\sigma^-$.
The general solution then takes the following form
\begin{gather}
\label{sols}
f(\alpha)=L \log({\sqrt{L^2-4\alpha^2}+\sqrt{\gamma^2-4\alpha^2}})+\gamma\log{\alpha\over \gamma\sqrt{L^2-4\alpha^2}
+L\sqrt{\gamma^2-4\alpha^2}}+C\;,
\end{gather}
where $C$ and $L$ are the integration constants. The constant $C$ can be, so that the field $X$
remains real.  To get an insight into the physics of these solutions it is instructive to inspect
the induced metric
 \[
 g_{ab}=\eta_{ab}+\ell_s^2\d_a X\d_b X\;,
 \]
 which determines the propagation of small perturbations. One observes that this family of solutions
 splits into two disconnected physical branches, $\mathcal{A}$ and $\mathcal{B}$, where the field
 takes real values.  The $\mathcal{A}$-branch covers the region
 \[
 4\alpha^2>L^2\,,\;\gamma^2\;,
 \]
 where one can introduce coordinates $(\rho,\lambda)$ defined as
 \[
 \sigma^\pm=\rho e^{\pm\lambda}\;.
 \]
 In these coordinates the induced metric becomes
 \[
 ds_{\cal A}^2={16\rho^4-L^2\gamma^2\over \rho^2(4\rho^2-L^2)}d\rho^2+2{L\gamma\sqrt{4\rho^2-\gamma^2}\over\rho\sqrt{4\rho^2-L^2}}d\rho d\lambda-
 (4\rho^2-\gamma^2)d\lambda^2
 \]
 By making a shift $\lambda\to\lambda+f(\rho)$ we can get rid of the off-diagonal term in the
 metric, so that it takes the Friedmann--Robertson--Walker form
 \be
 \label{FRW}
 ds_{\cal A}^2={16\rho^2d\rho^2\over 4\rho^2-L^2}-(4\rho^2-\gamma^2)d\lambda^2=d\tau^2-(\tau^2+L^2-\gamma^2)d\lambda^2\;,
 \ee
 where at the last step we introduced a new time coordinate $\tau=\sqrt{4\rho^2-L^2}$.

The  $\mathcal{B}$-branch of solutions covers the region
\[
4\alpha^2<L^2\,,\;\gamma^2\;.
 \]
One can repeat all the same steps here, and obtains the metric of the form
\[
 ds_{\cal B}^2=(\gamma^2-L^2+r^2)d\lambda^2-dr^2\;,
  \]
  where $r=\sqrt{L^2-4\rho^2}$.

We conclude that the $\mathcal{A}$-branch describes a cosmological solution and the
$\mathcal{B}$-branch corresponds to a static space-time. For $\gamma^2<L^2$ the cosmology is
non-singular, while the static solution exhibits a curvature singularity. For larger values of
$\gamma$ the singularity moves on the cosmological branch. This does not appear
very different from what one finds for a conventional string. The solutions take the same form there
with the role of time and space interchanged. There, for $\gamma^2<L^2$ one finds a smooth static
geometry and for $\gamma^2>L^2$ a non-singular bouncing cosmology.
\section{Discussion}
To summarize, it is fair to admit that our results are inconclusive at this point. We did not manage
to identify a clean pathology associated with a superluminal sign. Definitely, this theory allows
one to calculate a smaller set of observables than a conventional renormalizable quantum field
theory, but the same holds also for the conventional string world-sheet theory. Nevertheless, the
latter definitely gives rise to a rich and healthy system. The verdict for the former is not out
yet.  Both theories exhibit a number of gravitational features.  Gravitational theories are expected
to be less predictive than quantum field theories by not allowing one to calculate local off-shell
observables. Perhaps the most interesting lesson from the construction presented here, is that it
raises the question where is the proper place to stop on a slippery road between conventional UV
complete quantum field theories and non-renormalizable effective theories. The former allow sharp
prediction of both on-shell and off-shell observables.  The latter do not allow sharp predictions at
all in the mathematical sense, but often are quite adequate from the practical point of view.
Gravitational theories live in the middle.
\section{Acknowledgements}

We thank Raphael Flauger, Victor Gorbenko and  Sergey Sibiryakov for numerous helpful discussions.
This work is partially supported by the NSF grants PHY-1068438 and PHY-1316452. 
\bibliographystyle{utphys}
\bibliography{superlum}

\providecommand{\noopsort}[1]{}\providecommand{\singleletter}[1]{#1}%
\providecommand{\href}[2]{#2}\begingroup\raggedright\begin{thebibliography}{10}

\bibitem{Dvali:2000hr}
G.~Dvali, G.~Gabadadze, and M.~Porrati, ``{4-D gravity on a brane in 5-D
  Minkowski space},'' {\em Phys.Lett.} {\bf B485} (2000) 208--214,
\href{http://www.arXiv.org/abs/hep-th/0005016}{{\tt hep-th/0005016}}.

\bibitem{ArkaniHamed:2003uy}
N.~Arkani-Hamed, H.-C. Cheng, M.~A. Luty, and S.~Mukohyama, ``{Ghost
  condensation and a consistent infrared modification of gravity},'' {\em JHEP}
  {\bf 0405} (2004) 074,
\href{http://www.arXiv.org/abs/hep-th/0312099}{{\tt hep-th/0312099}}.

\bibitem{Dubovsky:2004sg}
S.~Dubovsky, ``{Phases of massive gravity},'' {\em JHEP} {\bf 0410} (2004) 076,
\href{http://www.arXiv.org/abs/hep-th/0409124}{{\tt hep-th/0409124}}.

\bibitem{Nicolis:2008in}
A.~Nicolis, R.~Rattazzi, and E.~Trincherini, ``{The Galileon as a local
  modification of gravity},'' {\em Phys.Rev.} {\bf D79} (2009) 064036,
\href{http://www.arXiv.org/abs/0811.2197}{{\tt 0811.2197}}.

\bibitem{deRham:2010kj}
C.~de~Rham, G.~Gabadadze, and A.~J. Tolley, ``{Resummation of Massive
  Gravity},'' {\em Phys.Rev.Lett.} {\bf 106} (2011) 231101,
\href{http://www.arXiv.org/abs/1011.1232}{{\tt 1011.1232}}.

\bibitem{Dvali:2002vf}
G.~Dvali, A.~Gruzinov, and M.~Zaldarriaga, ``{The Accelerated universe and the
  moon},'' {\em Phys.Rev.} {\bf D68} (2003) 024012,
\href{http://www.arXiv.org/abs/hep-ph/0212069}{{\tt hep-ph/0212069}}.

\bibitem{Lue:2002sw}
A.~Lue and G.~Starkman, ``{Gravitational leakage into extra dimensions: Probing
  dark energy using local gravity},'' {\em Phys.Rev.} {\bf D67} (2003) 064002,
\href{http://www.arXiv.org/abs/astro-ph/0212083}{{\tt astro-ph/0212083}}.

\bibitem{Dubovsky:2004ud}
S.~Dubovsky, P.~Tinyakov, and I.~Tkachev, ``{Massive graviton as a testable
  cold dark matter candidate},'' {\em Phys.Rev.Lett.} {\bf 94} (2005) 181102,
\href{http://www.arXiv.org/abs/hep-th/0411158}{{\tt hep-th/0411158}}.

\bibitem{Dubovsky:2007zi}
S.~Dubovsky, P.~Tinyakov, and M.~Zaldarriaga, ``{Bumpy black holes from
  spontaneous Lorentz violation},'' {\em JHEP} {\bf 0711} (2007) 083,
\href{http://www.arXiv.org/abs/0706.0288}{{\tt 0706.0288}}.

\bibitem{Dubovsky:2009xk}
S.~Dubovsky, R.~Flauger, A.~Starobinsky, and I.~Tkachev, ``{Signatures of a
  Graviton Mass in the Cosmic Microwave Background},'' {\em Phys.Rev.} {\bf
  D81} (2010) 023523,
\href{http://www.arXiv.org/abs/0907.1658}{{\tt 0907.1658}}.

\bibitem{Hui:2012jb}
L.~Hui and A.~Nicolis, ``{Proposal for an Observational Test of the Vainshtein
  Mechanism},'' {\em Phys.Rev.Lett.} {\bf 109} (2012) 051304,
\href{http://www.arXiv.org/abs/1201.1508}{{\tt 1201.1508}}.

\bibitem{Adams:2006sv}
A.~Adams, N.~Arkani-Hamed, S.~Dubovsky, A.~Nicolis, and R.~Rattazzi,
  ``{Causality, analyticity and an IR obstruction to UV completion},'' {\em
  JHEP} {\bf 0610} (2006) 014,
\href{http://www.arXiv.org/abs/hep-th/0602178}{{\tt hep-th/0602178}}.

\bibitem{Dubovsky:2006vk}
S.~Dubovsky and S.~Sibiryakov, ``{Spontaneous breaking of Lorentz invariance,
  black holes and perpetuum mobile of the 2nd kind},'' {\em Phys.Lett.} {\bf
  B638} (2006) 509--514,
\href{http://www.arXiv.org/abs/hep-th/0603158}{{\tt hep-th/0603158}}.

\bibitem{Dubovsky:2008bd}
S.~Dubovsky and S.~Sibiryakov, ``{Superluminal Travel Made Possible (in two
  dimensions)},'' {\em JHEP} {\bf 0812} (2008) 092,
\href{http://www.arXiv.org/abs/0806.1534}{{\tt 0806.1534}}.

\bibitem{Dubovsky:2011ii}
S.~Dubovsky and V.~Gorbenko, ``{Superluminal Travel, UV/IR Mixing and
  Turbulence in the Lineland},'' {\em Phys.Rev.} {\bf D84} (2011) 105039,
\href{http://www.arXiv.org/abs/1108.2891}{{\tt 1108.2891}}.

\bibitem{Dubovsky:2012sh}
S.~Dubovsky, R.~Flauger, and V.~Gorbenko, ``{Effective String Theory
  Revisited},'' {\em JHEP} {\bf 1209} (2012) 044,
\href{http://www.arXiv.org/abs/1203.1054}{{\tt 1203.1054}}.

\bibitem{Dubovsky:2012wk}
S.~Dubovsky, R.~Flauger, and V.~Gorbenko, ``{Solving the Simplest Theory of
  Quantum Gravity},'' {\em JHEP} {\bf 1209} (2012) 133,
\href{http://www.arXiv.org/abs/1205.6805}{{\tt 1205.6805}}.

\bibitem{Goddard:1973qh}
P.~Goddard, J.~Goldstone, C.~Rebbi, and C.~B. Thorn, ``{Quantum dynamics of a
  massless relativistic string},'' {\em Nucl.Phys.} {\bf B56} (1973)
109--135.

\bibitem{Zamolodchikov:1989cf}
A.~Zamolodchikov, ``{Thermodynamic Bethe Ansatz in relativistic models. Scaling
  three state Potts and Lee-Yang models},'' {\em Nucl.Phys.} {\bf B342} (1990)
695--720.

\bibitem{Dorey:1996re}
P.~Dorey and R.~Tateo, ``{Excited states by analytic continuation of TBA
  equations},'' {\em Nucl.Phys.} {\bf B482} (1996) 639--659,
\href{http://www.arXiv.org/abs/hep-th/9607167}{{\tt hep-th/9607167}}.

\bibitem{Mezincescu:2010yp}
L.~Mezincescu and P.~K. Townsend, ``{Anyons from Strings},'' {\em
  Phys.Rev.Lett.} {\bf 105} (2010) 191601,
\href{http://www.arXiv.org/abs/1008.2334}{{\tt 1008.2334}}.

\bibitem{Polchinski:1991ax}
J.~Polchinski and A.~Strominger, ``{Effective string theory},'' {\em
  Phys.Rev.Lett.} {\bf 67} (1991) 1681--1684.

\end{thebibliography}\endgroup

\end{document}